\documentclass[a4paper,12pt]{article}

\usepackage{fullpage}
\usepackage[top=3cm, bottom=3cm, left=3cm, right=3cm]{geometry}
\usepackage{natbib}
\usepackage{amssymb}
\usepackage{amsmath}
\usepackage{amsthm}
\usepackage{amsbsy}
\usepackage{amsfonts}
\usepackage{float}
\usepackage{setspace}
\usepackage[usenames,dvipsnames]{xcolor}
\usepackage{graphicx}
\usepackage{titling}
\usepackage{bbm}
\usepackage{dsfont}
\usepackage{color}
\usepackage[dvips]{epsfig}
\usepackage{verbatim}
\usepackage{nomencl}
\usepackage{multirow}
\usepackage{caption}
\usepackage{subfig}
\usepackage{fancyref}

\newcommand{\bm}[1]{{\boldsymbol {#1}}}

\begin{document}

\title{Accounting for parameter uncertainty in two-stage designs for
  Phase II dose-response studies}

\author{Emma McCallum$^1$ and Bj\"{o}rn Bornkamp$^{2}$\\
\small{$^1$ MRC Biostatistics Unit, Cambridge, United Kingdom}\\
\small{$^2$ Novartis Pharma AG, Basel, Switzerland}}

\date{}

\maketitle

\begin{abstract}
  In this paper we consider two-stage adaptive dose-response study
  designs, where the study design is changed at an interim analysis
  based on the information collected so far. In a simulation study,
  two approaches will be compared for these type of designs; (i)
  updating the study design by calculating the maximum likelihood
  estimate for the dose-response model parameters and then calculating
  the design for the second stage that is locally optimal for this
  estimate, and (ii) using the complete posterior distribution of the
  model parameter at interim to calculate a Bayesian optimal design
  (i.e. taking into account parameter uncertainty). In particular, for
  an early interim analysis respecting parameter uncertainty seems
  more adequate, on the other hand for a Bayesian approach dependency
  on the prior is expected and an adequately thought-through prior is
  required. A computationally efficient method is proposed for
  calculating the Bayesian design at interim based on approximating
  the full posterior sample using k-means clustering.  The sigmoid
  Emax dose-response model and the D-optimality criterion will be used
  in this paper.
\end{abstract}

\textbf{Keywords:} adaptive dose-finding, Bayesian optimal design,
D-optimality, k-means clustering, Hill equation, sigmoid Emax

\setstretch{1.5}

\section{Introduction}

In the past few years there has been substantial interest in adaptive
dose-finding studies, see for example, the white papers of the PhRMA
Working Group for Adaptive Dose-Ranging Designs \citep{Bornkamp07,
  Dragalin10}. Other references include \citet{Grieve05, Dragalin07,
  Miller07, Berry10, Bornkamp11, Jones11} among many others. As fixed
designs are more often used in clinical practice, one obvious question
is to characterize situations when an adaptive design will outperform
a non-adaptive design from a statistical efficiency perspective. In
practice, of course also non-statistical (or alternative statistical)
considerations will play a role at the trial design stage, when one
decides for or against using an adaptive design, but we will focus on
statistical estimation efficiency in this paper. 

To that end, \citet{Dette13} investigated two-stage designs, where an interim
analysis is performed and the maximum likelihood (ML) estimate of the
model parameters is used to calculate a locally optimal design for the
second stage of the trial. Based on analytical approximations they
identified key factors, when adaptive designs are beneficial,
including signal to noise ratio (the larger the better for adaptive
designs), timing of the interim analysis (if the interim estimate is
too early the interim estimate might be too noisy to produce a good
design, whereas if the interim analysis is too late there might be not enough
patients left to allocate into the trial for the calculated optimal
design), and adequacy of the starting design (if an adequate starting
design is used, there might not be much to improve).

A problem of the ML approach discussed in \citet{Dette13} is that the
study design is optimized based on the ML point estimate at the
interim. By using only a point estimate, the uncertainty in the
parameter estimate is ignored and potentially plausible parameter
values discarded, ultimately leading to designs optimized for the
``wrong'' parameter values. Uncertainty can be accounted for at the
interim analysis by adopting a Bayesian approach with prior
distributions for the model parameters specified at the beginning of
the trial and the posterior distribution based on the first stage data
being utilized at the interim analysis to update the study design.

The primary aim of this work is to investigate whether there are any
gains to be made by accounting for parameter uncertainty at the
interim by using Bayesian methods or whether ML updating of the
optimal design is preferable. As analytical considerations such as
those discussed in \citet{Dette13} are infeasible for the Bayesian
designs considered here, we conducted a large-scale simulation study
in the realistically relevant settings of the PhRMA working group. As
we concentrate on parameter uncertainty in this paper and not model
uncertainty, we only use the sigmoid Emax model \citep{MacDougall06}:
It is commonly used for modeling the dose-response relationship and
its main assumption is monotonicity which is a desirable feature for
modeling many clinical trial dose-response curves \citep{Kirby10}.  In
addition the sigmoid Emax model has been found to give a good
approximation to a number of other possible dose-response models as
well \citep{Thomas06}. The design criterion used will be
D-optimality. The type of adaptation under consideration involves
changing the randomisation ratios to existing dose-levels based on
efficacy.

\section{Methodology}
\subsection{Statistical model}
Let $Y$ be a clinical outcome observed at a dose level $x \in [0,D]$
where $0$ and $D$ are the placebo and maximum dose, respectively.  A
parallel group design is used where several active dose levels plus a
placebo are considered, with patients being randomized to one of the
treatments. There are $q$ discrete doses available, with $q$ being a
fixed number and the grid of available doses being denoted
$\mathbf{x}=(x_1, \ldots,x_q)$.
\\*
\par
Consider the following regression model
\begin{equation}\label{eqn:model}
Y_{ij}=\eta (x_i,\boldsymbol{\theta})+\epsilon_{ij}, \quad \epsilon_{ij} \sim N(0, \sigma^2)
\end{equation}
where $Y_{ij}$ denotes the response of patient $j$ at dose $x_i$
($i=1,\ldots,q$,$j=1,\ldots,n_i$) and $\epsilon_{ij}$ is the residual
error, which is assumed to be independent and normally distributed
with common variance $\sigma^2$. The true, but unknown, $r$-parameter
dose-response model is denoted by $\eta (x_i,\boldsymbol{\theta})$
with parameter vector
$\boldsymbol{\theta}=(\theta_1,\ldots,\theta_r)$.
The total sample size is $\sum_{i=1}^{q}n_i=N$ where $n_i$ denotes the number of patients
assigned to dose $x_i$.

\subsubsection{Sigmoid Emax model}
The four parameter sigmoid Emax model is defined as
\begin{equation}\label{eqn:emax}
\eta(x_i,\boldsymbol{\theta})=\theta_1+\theta_2\frac{x_i^{\theta_4}}{\theta_3^{\theta_4}+x_i^{\theta_4}}
\end{equation}
where $\theta_1$ denotes the placebo effect, $\theta_2$ is the
asymptotic maximum treatment effect (often called Emax),
$\theta_3$ is the dose that gives half of the maximum treatment effect
(often called $ED_{50}$), and $\theta_4$ is the so-called Hill
parameter that determines the steepness of the curve.

The gradient of the sigmoid Emax model,
$\mathbf{g}(x,\boldsymbol{\theta})$, is given as
\begin{equation}\label{eqn:gradient}
 \mathbf{g}(x,\boldsymbol{\theta}) = \left(1, \frac{1}{1+(\theta_3/x)^{\theta_4}}, \frac{-\theta_4 \theta_2 \theta_3^{(\theta_4-1)}}{x^{\theta_4}(1+(\theta_3/x)^{\theta_4})^2}, \frac{-\theta_2\log(\theta_3/x)(\theta_3/x)^{\theta_4}}{(1+(\theta_3/x)^{\theta_4})^2} \right)
 \end{equation}

 Here there are discrete dose levels and so the Fisher information
 matrix for the statistical model given in (\ref{eqn:model}) for a
 sigmoid Emax model is defined as
\begin{equation}\label{eqn:FIM}
\boldsymbol{M} (\mathbf{x}, \mathbf{w}, \boldsymbol{\theta}) = \sum_{i=1}^{q} w_i \mathbf{g}(x_i, \boldsymbol{\theta})^T\mathbf{g}(x_i, \boldsymbol{\theta})
\end{equation}
where $w_i=n_i/N$ is the proportion of patients allocated to dose
$x_i$ and $\mathbf{w}=(w_1,\ldots,w_q)$. If a dose from the grid of
available doses ${x_1, \ldots,x_q}$ is not allocated any patients then
$w_i=0$. Note that the Fisher information matrix for the nonlinear
sigmoid Emax model (Equation \ref{eqn:FIM}) is dependent on the
unknown true parameter values $\boldsymbol{\theta}$ and the
experimental design $\xi=(\mathbf{x}, \mathbf{w})$.

\subsection{Two-stage D-optimal design}
A fixed (or non-adaptive) design is a design in which observations are
taken at pre-specified doses. In a two-stage adaptive design, the data
accrued in the first stage is used to determine the design for the
second stage, as follows;
\begin{enumerate}
\item Stage 1: $N^{(1)}$ observations at starting design $\xi^{(1)}$.
\item Interim updating: fit the sigmoid Emax model to the first stage
  data and use the obtained information about the model parameters
  $\boldsymbol{\theta}$ to calculate the optimal design for the second
  stage, $\xi^{(2)}$.
\item Stage 2: $N^{(2)}=N-N^{(1)}$ observations based on updated
  design $\xi^{(2)}$.
\end{enumerate}
Step 2 in the procedure above will be different for the ML-based
updating and the Bayesian updating. In this paper, the entire
dose-response curve is of interest and so the $D$-optimal design is
calculated at the interim analysis. A design is $D$-optimal if it
maximizes the determinant of the Fisher information matrix or,
equivalently, minimizes the generalized variance of the parameter
estimates. Such an optimal design minimizes the volume of the
confidence ellipsoid of the parameters, and minimizes the maximum
predicted variance around the dose-response curve
\citep[Ch.10,11]{Atkinson07}.

\subsection{Interim updating}
For both types of updating the same grid of pre-specified discrete
doses are available at the interim and so the weights are optimised
whilst the doses remain fixed.  A continuous optimal design is
calculated and then rounded to a fixed number of patients so that the
total number of patients $N$ is achieved \citep[Chapter
12]{Pukelsheim06}.

\subsubsection{Maximum Likelihood updating}
\label{sec:maxim-likel-updat}

At the interim, the sigmoid Emax model is fitted to the data from the
first stage using the \verb="fitMod"= function from the
\verb=DoseFinding= R package \citep{Bornkamp13} which implements
nonlinear least squares regression. Bounds will be used for the
nonlinear parameters $\theta_3$ and $\theta_4$; this is a sufficient
condition for existence of the ML estimate, see \citet{Jennrich1969}.
The bounds $[0.001, 1.5D]$ for the $\theta_3$ parameter and $[0.5,
10]$ for the $\theta_4$ will be used (here $D$ is the maximum used
dose). Both boundaries are chosen relatively wide to allow essentially
all shapes of the underlying sigmoid Emax shape in the considered
dose-range $[0,D]$.

After the ML estimate has been calculated the estimates are then input
to the Fisher information matrix (Equation \ref{eqn:FIM}). The optimal
design for the second stage, $\xi^{(2)}$, is found by maximising the
vector of weights, $\mathbf{w}$, in the following expression:

\begin{equation}
\xi^{(2)}= \arg \max_{\mathbf{w}}|\mathbf{M} (\mathbf{x}, \mathbf{w}, \boldsymbol{\hat{\theta}}^{(1)})| = \arg \max_{\mathbf{w}} \det \left[\sum_{i=1}^{q} w_i \mathbf{g}(x_i, \boldsymbol{\hat{\theta}}^{(1)})^T\mathbf{g}(x_i, \boldsymbol{\hat{\theta}}^{(1)}) \right]
\end{equation}

This optimisation is carried out using the \verb="optDesign"= function
in the DoseFinding package in \verb=R=, which performs a nonlinear
optimization using an augmented Lagrange method. The remaining
$N^{(2)}$ patients are then assigned to the optimal design,
$\xi^{(2)}$, for the second stage.

\subsubsection{Bayesian updating}
The conceptual framework behind the Bayesian optimal design is the
assumption that the information of the parameters of interest and
their uncertainty can be adequately captured in the probability
distribution. This probability density averages out the parameter values and the
criterion is no longer dependent on the parameters. The Bayesian
D-optimality criterion is defined in \citet{Atkinson07} as
\begin{equation}\label{eqn:BayesianDopt}
\Psi_B(\xi)=\mathrm{E}_{\boldsymbol{\theta}} \left[\log |\mathbf{M}(\xi,\boldsymbol{\theta})|\right]=\int_{\theta} \log |\mathbf{M}(\xi,\boldsymbol{\theta})|p(\boldsymbol{\theta}) d\boldsymbol{\theta}
\end{equation}
where $\mathbf{M}(\xi,\boldsymbol{\theta})$ is the Fisher information
matrix and $p(\boldsymbol{\theta})$ is the density capturing the
information on $\boldsymbol{\theta}$.  This type of design is more
robust than a locally optimal design in the sense that the performance
will be adequate for parameter values that have relevant probability
mass, whereas for a locally optimal design one only considers a single
parameter value. In many situations $p(\boldsymbol{\theta})$ might be
a prior distribution; in this paper the distribution used in the
design criterion will be the posterior distribution
$p(\boldsymbol{\theta}|\bm y^{(1)})$ based on the data observed in the
first stage of the trial. Note that this distribution will contain
less information (\textit{e.g.} it will have a larger posterior
variance) if an early interim analysis is performed. When the
posterior is calculated on a late interim the density will be more
peaked.  The ``robustness'' of the design criterion hence
automatically adapts to the amount of information available. The ML
based approach in contrast always uses the point estimate without
acknowledging the uncertainty associated with it.

For the sample size and signal-to-noise ratio of interest in
dose-finding studies there will be a dependence on the prior
distribution specified at the beginning of the trial
(\citet{Temple12}; \citet{Bornkamp12}), in particular if the interim
analysis is performed early in the trial. For the purpose of generic
simulations, outside a concrete real example, it is difficult to come
up with examples of prior information one might have, so the focus in
this paper will be on the situation of little prior information. It is
clear, that a fully Bayesian approach will be better than the
maximum-likelihood approach if relevant historical prior information
is available, so the situation considered in the simulations might be
considered conservative for the performance of the Bayesian approach.

\paragraph{The functional uniform prior distribution}
\mbox{}\\
One potential weakly informative prior distribution is the Jeffreys
prior distribution, as for example discussed in
\citep[p.63]{Gelman03}. Here one uses
\begin{equation}
  p(\boldsymbol{\theta}) \propto |\mathbf{M}(\mathbf{x}, \mathbf{w},\boldsymbol{\theta})|^{1/2},
\end{equation}
where $\boldsymbol{M} (\mathbf{x}, \mathbf{w}, \boldsymbol{\theta})$
is the Fisher information matrix defined in Equation
(\ref{eqn:FIM}). So the Jeffreys prior depends on the Fisher
information matrix, which in turn depends on the \textit{observed}
design $(\mathbf{x}, \mathbf{w})$. However, in a two-stage trial the
design at the interim will be different to the design at study end, so
that the Jeffreys prior will be different at interim and study end,
which is why the prior violates the likelihood principle
\citep{Ghosh06}.

Functional uniform priors \citep{Bornkamp12, Bornkamp14} strive to
derive a distribution that is uniformly distributed on the potential
different shapes of the underlying nonlinear model function. These
priors are invariant with respect to parameterization of the model
function and typically result in rather non-uniform prior
distributions on the parameter scale. When using the $L_2$ distance at
the grid of available doses ${x_1, \ldots,x_q}$ with equal weight to
define functional uniformity, one obtains a functional uniform
distribution that is essentially a modified version of Jeffreys prior
\citep{Bornkamp12}. Instead of the actual observed design one uses the
$\mathbf{x}$ values and weights for $\mathbf{w}$ used in the $L_2$
distance. Here we will use the grid $\mathbf{x}$ of all doses
available and equal weights $\mathbf{w}$. This prior will then no
longer violate the likelihood principle (it is completely specified
before trial start) and has the clear interpretation of being
uniformly distributed on the different shapes available in the chosen
distance metric. The same parameter bounds as for ML estimation will
be used for the Bayesian approach. This is the prior used in this
paper.

\paragraph{Numerical calculations to update Bayesian designs}
\mbox{}\\

Two numerically challenging tasks are involved in calculation of the
updated design at the interim analysis based on the Bayesian
approach; (i) one needs to perform the integration in Equation
(\ref{eqn:BayesianDopt}) to evaluate the design criterion once, and (ii)
one needs to optimize the design criterion with respect to the design,
requiring multiple evaluations of Equation (\ref{eqn:BayesianDopt})
for different candidate designs.

One approach to perform the integration is to use Monte Carlo. That
means at the interim analysis JAGS will be used to create a Markov
chain Monte Carlo sample of size 10000 from the posterior
distribution. The integral can then be estimated as the corresponding
arithmetic mean. So for 10000 samples from the posterior distribution
this requires calculation of 10000 evaluations of the determinant of
the Fisher information to evaluate the criterion for one design. This
can become time consuming, in particular because evaluation of the
design criterion is embedded in the numerical optimization. In context
of a simulation study this approach would be computationally
prohibitive.

A more efficient alternative to pure Monte Carlo is to use k-means
clustering to approximate the full posterior distribution for the
parameters by a weighted discrete distribution. The k-means cluster
algorithm partitions $n$ observations into $k$ clusters so as to
minimise the within-cluster sum of squares \citep{Wu12}. So instead of
using all posterior samples, each weighted equally, a weighted mean is
used, based on the identified cluster centers (and corresponding
cluster weights), so that the determinant of the Fisher information
needs to be evaluated less often for one evaluation of the design
criterion. The \citet{Hartigan79} k-means clustering algorithm
implemented in \verb=R= in the function \verb=kmeans= will be used.

Theoretically for a large number of cluster centers (and a large
posterior sample), the k-means approximation of the integral should be
close to the correct integral, in Appendix \ref{apdx:kmeans} we
outline some heuristic mathematical arguments that support this. In
practice it is unclear how many cluster centers will provide an
adequate approximation of the integral and so a small simulation study
was conducted, where $k=10$ cluster centers were used. Actual
calculation of the optimal design was performed with the
\verb="optDesign"= function, where the cluster centers were used for
the model parameters and the corresponding cluster weights specified.
The cluster weights correspond to the proportion of posterior samples
associated with each of the cluster centres.

We used one of the simulation scenarios studied in the simulation
study described later to evaluate the loss of efficiency when using
the k-means approximation instead of the full posterior, see Appendix 
\ref{apdx:scenario} for the exact simulation scenario. The relative
performance is defined as

\begin{equation}
\mathrm{Relative Performance}=\frac{\Psi_B(\xi^{*}_{\mathrm{km}})}{\Psi_B(\xi^{*}_{\mathrm{fp}})}
\end{equation}
where $\xi^{*}_{\mathrm{km}}$ is the optimal design calculated using
the k-means approximation, $\xi^{*}_{\mathrm{fp}}$ is the optimal
design calculated using the full posterior sample and $\Psi_B$ the
Bayes design criterion from Equation (\ref{eqn:BayesianDopt}).

\begin{figure}[h]
\begin{center}
\includegraphics[scale=0.33]{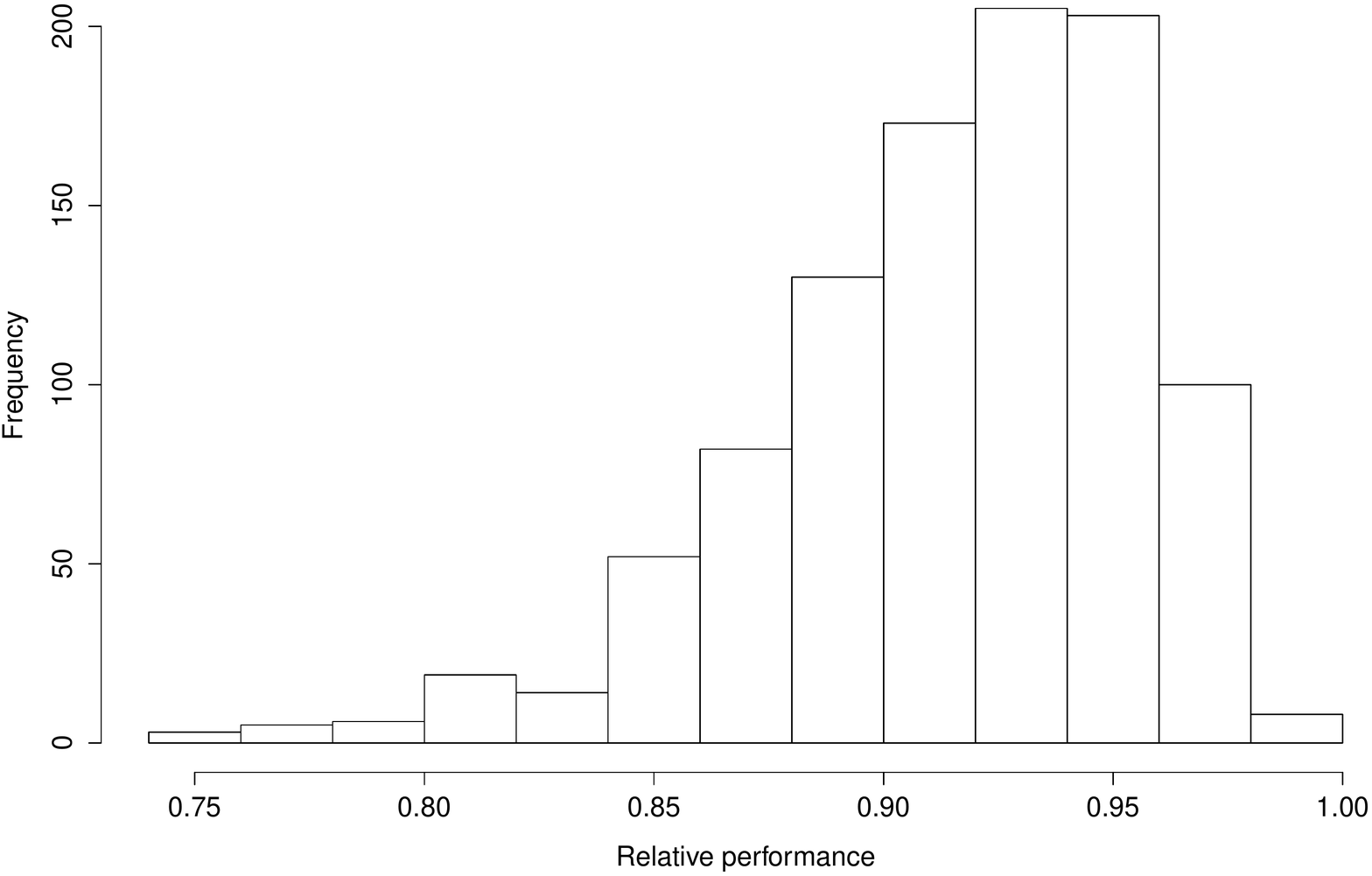}
\end{center}
\caption{Relative performance of k-means approximation to full posterior}
\label{eff_kmeans15}
\end{figure}

Figure \ref{eff_kmeans15} shows one example of the distribution of
efficiencies for 1000 datasets simulated using a sigmoid Emax model
with $15$ observations. The sample size was chosen to be small, as in
these situations it is assumed that the posterior has the most complex
shape, and is most difficult to approximate.

The k-means approximation for the full posterior is fairly good with
an average efficiency of $0.915$ for this scenario. The approximation
improves as the sample size in the first stage of the trial increases
with an average efficiency of $0.961$ and $0.976$, for sample sizes 60
and 150, respectively.

\begin{figure}[h]
\centering
\subfloat[K-means approximation]{\includegraphics[width=0.48\textwidth]{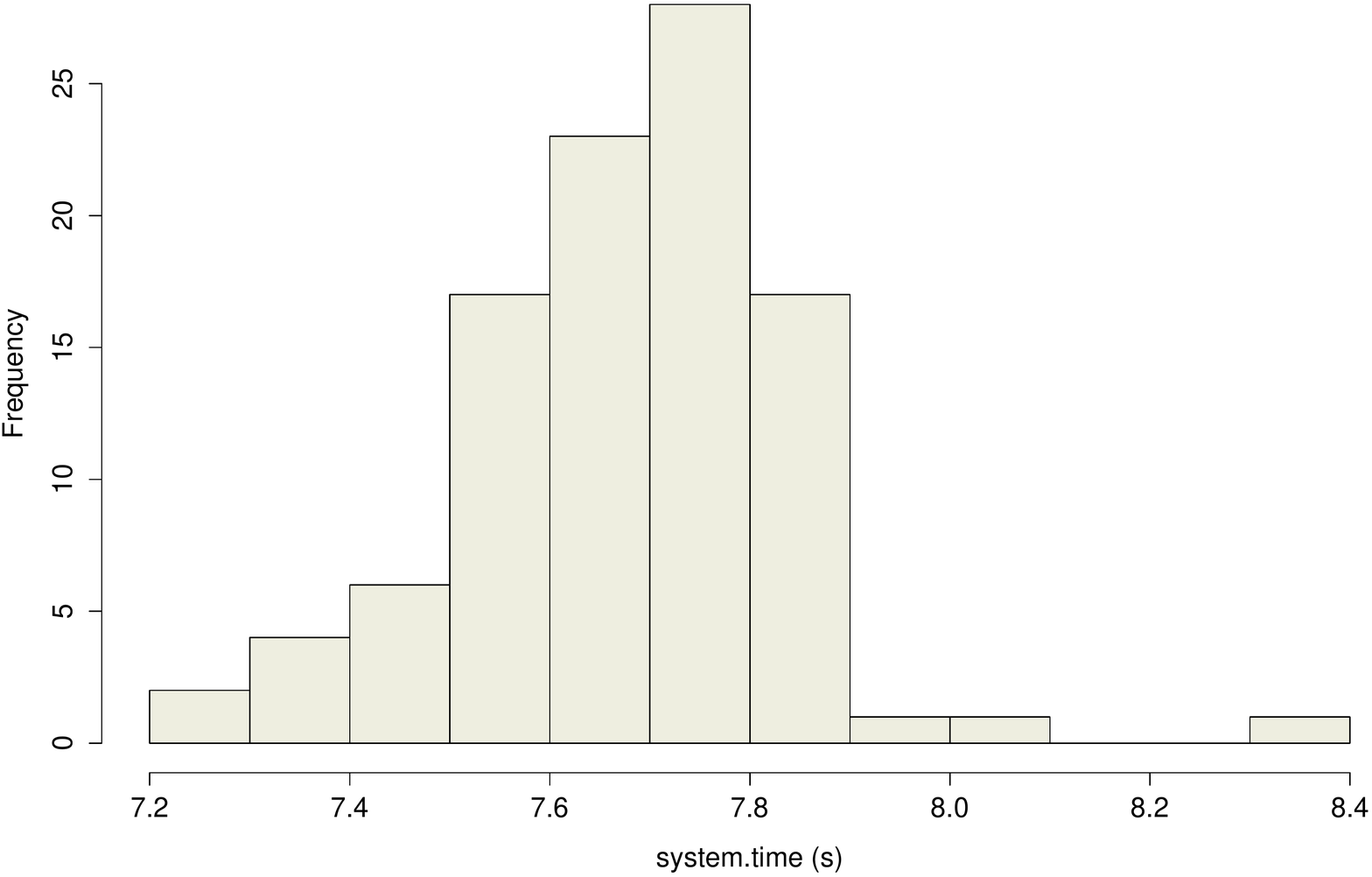}}
~
\subfloat[Full posterior distribution]{\includegraphics[width=0.48\textwidth]{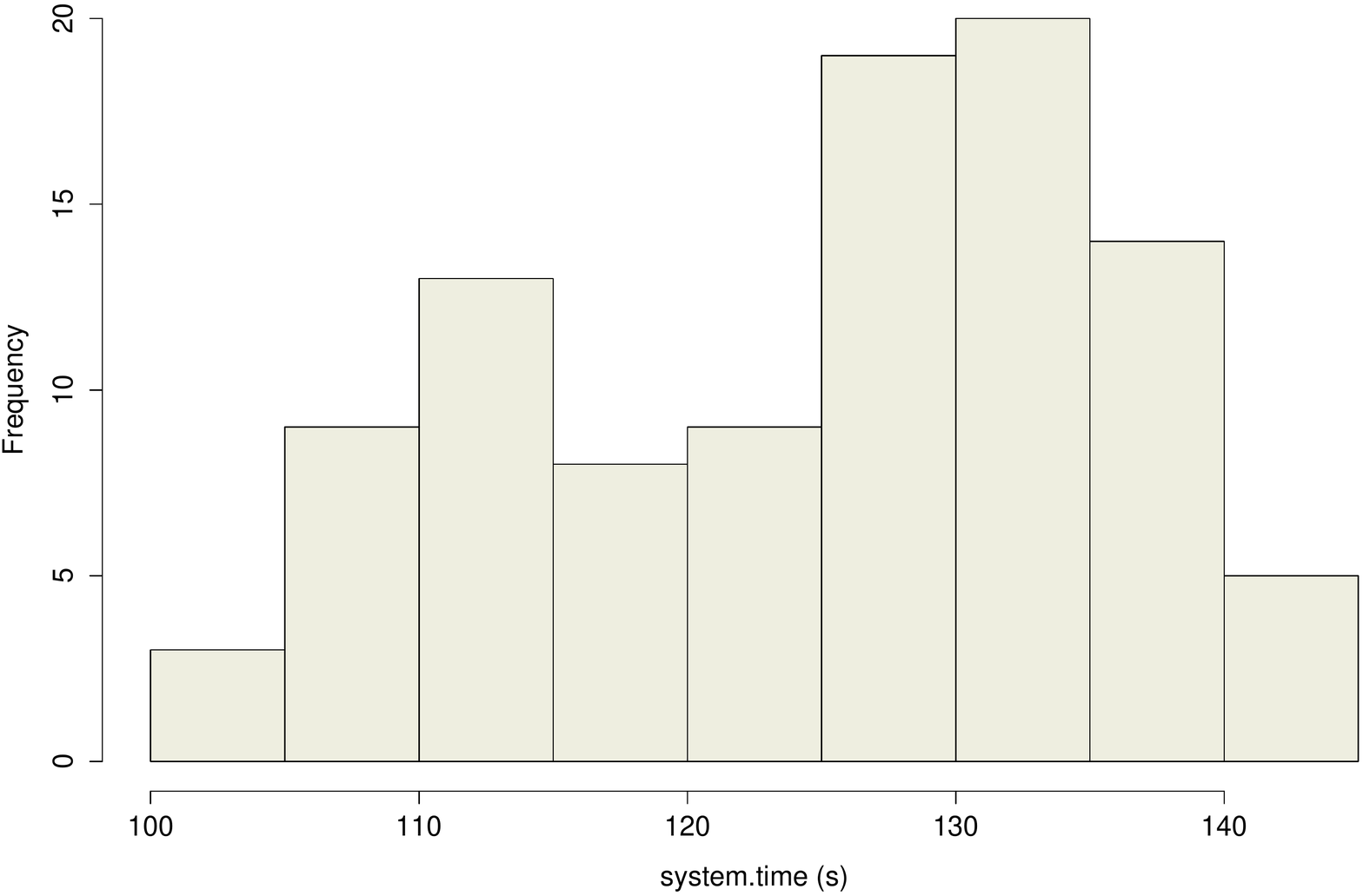}}
\caption{Distribution of running times for calculation of the second stage design}\label{fig:kmeans-time}
\end{figure}

In a scenario with a total sample size of 250 the k-means
approximation required on average 7.7 seconds to calculate the optimal
design, compared to 124.9 seconds for the calculation of optimal
design using the full posterior distribution (refer to Appendix 
\ref{apdx:scenario} for the simulation scenario). The distributions of the
running time for 100 simulations are shown in Figure
\ref{fig:kmeans-time}, when using (a) the k-means approximation and
(b) the full posterior distribution. So a more than 10-fold reduction
in computing time can be achieved by using this approximation.

\section{Simulations}

In total, 448 different scenarios were considered in the
simulations. Each scenario correspond to a different combination of
starting design, dose-response profile, total sample size, timing of
the interim analysis and type of interim updating. The scenarios
selected are similar to those conducted by the PhRMA Adaptive
Dose-Ranging Designs Working Group \citep{Bornkamp07}. In the PhRMA
Working Group paper, a comprehensive simulation study based on a
hypothetical neuropathic pain study was conducted to evaluate existing
adaptive dose-ranging methods. The primary endpoint considered was the
change from baseline at 6 weeks in a visual analog scale (VAS) of pain
and this is taken to be the primary endpoint in the following
simulation study. The VAS measurements are on a continuous scale from
0 (no pain) to 10 (highest pain). If the VAS measurement at the $k$th
week is denoted by $VAS_k$ with $k=0,1,\ldots,6$ and $k=0$
representing the baseline, then the primary endpoint is defined as
\begin{equation}
y=VAS_6-VAS_0
\end{equation}
Hence, negative values of $y$ indicate efficacy as there is a
reduction in neuropathic pain.

\subsection{Starting designs}
In order to investigate the impact of number and spacing of doses,
four different starting designs are considered in the simulations
(Table \ref{tab:designs}). For the first stage, patients are allocated
equally between the doses considered.

\begin{table}[h]
\centering
\renewcommand{\arraystretch}{1.5}
\begin{tabular}{|l|l|l|}
  \hline
  Design A & Five equally spaced doses & $\xi_A=(0, 2, 4, 6, 8)$ \\
  Design B & Five unequally spaced doses (mostly low doses) & $\xi_B=(0, 1, 2, 4, 8)$ \\
  Design C & Five unequally spaced doses (mostly high doses) & $\xi_C=(0,6,7,7.5,8)$ \\
  Design D & Nine equally spaced doses & $\xi_D=(0,1,2,3,4,5,6,7,8)$ \\
  \hline
\end{tabular}
\caption{Starting designs}
\label{tab:designs}
\end{table}

Designs A and B reflect what is most commonly seen in practice, in
particular, design B is an equal distribution on the $\log_2$
scale. Design C is more unusual but can occur if the dose range
considered is too narrow and the lowest dose is not low enough. Design
D is less likely to be seen in practice as it is unusual to see so
many dose levels utilized.

At the interim analysis it is assumed that 17 equally spaced doses can
be utilized ranging from $0$ to $8$ in $0.5$ increments:
$\mathbf{x}=(0,0.5,1,1.5,2,\ldots,7.5,8)$.

\subsection{Dose-response profiles}

\begin{figure}[h]
\begin{center}
\includegraphics[scale=0.6]{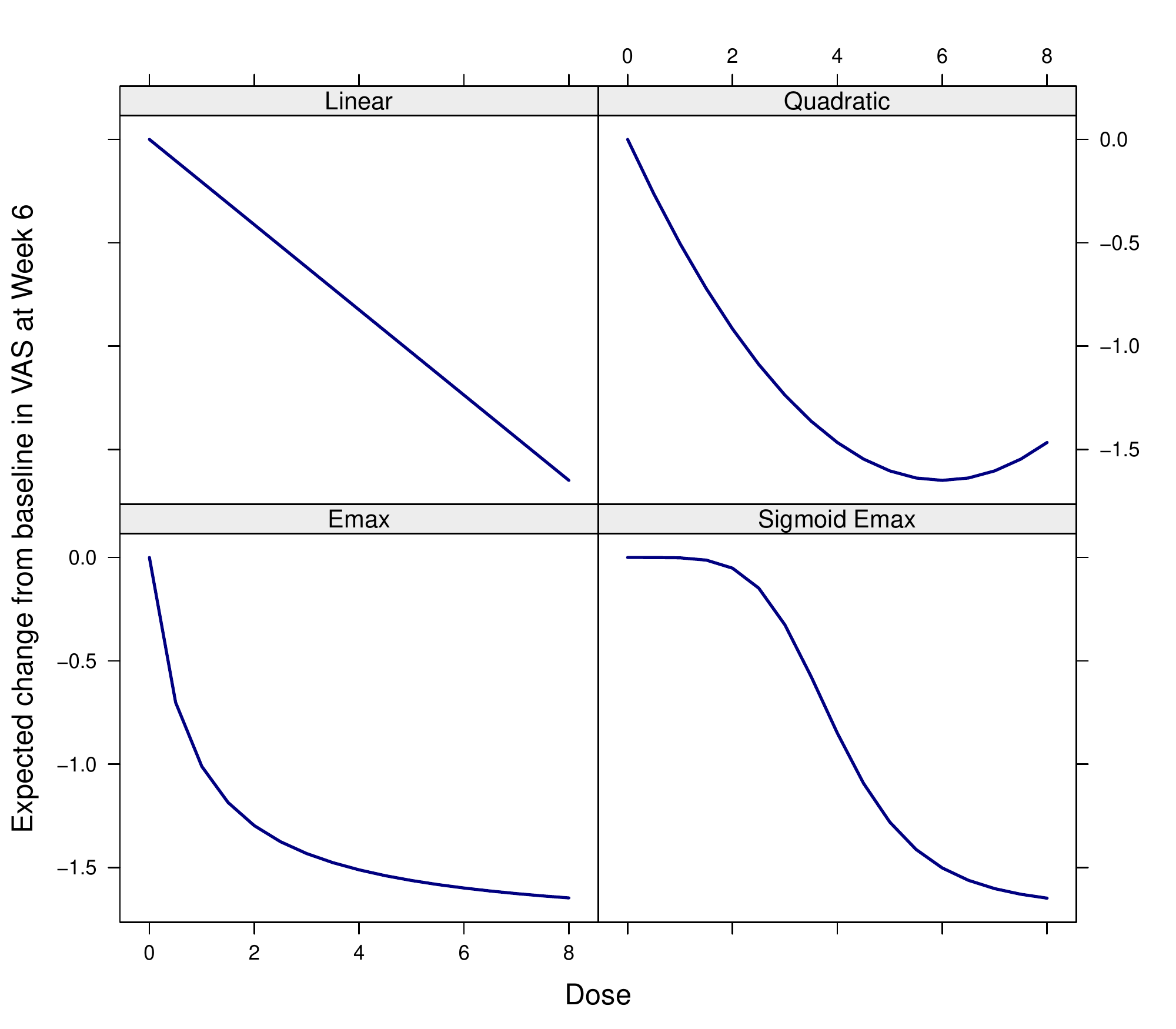}
\end{center}
\caption{Dose-response profiles}
\label{DRcurves}
\end{figure}

A total of four different dose-response profiles were used to simulate
the primary endpoint, which are considered to span a range of
dose-response profiles often observed in practice (Figure
\ref{DRcurves}). As in \citet{Bornkamp07}, for all models the placebo
effect was set to 0 and the maximum effect within the observed dose
range was set to $-1.65$ units.

\begin{table}[h]
\centering
\renewcommand{\arraystretch}{1.8}
\begin{tabular}{r l}
Linear: & $\eta(x,\boldsymbol{\theta})=-(1.65/8)x$ \\
Quadratic: & $\eta(x,\boldsymbol{\theta})=-(1.65/3)x+(1.65/36)x^2$ \\
Emax: & $\eta(x,\boldsymbol{\theta})=-1.81x/(0.79+x)$ \\
Sigmoid Emax: & $\eta(x,\boldsymbol{\theta})=-1.70x^5/(4^5+x^5)$ \\
\end{tabular}
\end{table}

The residual error is assumed to be independently normally distributed
with mean $0$ and variance $4.5$.

Note that despite the fact that the linear and quadratic model are
used here as true simulation scenarios, we will always only use the
sigmoid Emax model to update the design and analyse the simulated
studies. So for the linear and quadratic scenarios we investigate the
behaviour of the procedure under mis-specification of the
dose-response model.

\subsection{Sample size and timing of interim analysis}
Two total sample sizes are used in the simulations: 150 and 250
patients. These sample sizes are consistent with those commonly used
in neuropathic pain dose-finding studies \citep{Bornkamp07}. The
timing of the interim analysis was varied in order to consider the
impact of sample size in the first stage. Seven different timings for
the interim analysis were considered for each total sample size (Table
\ref{tab:timing}).

\begin{table}[H]
\centering
\renewcommand{\arraystretch}{1.8}
\begin{tabular}{|l|c|c|c|c|c|c|c|}
\hline
  &  \multicolumn{7}{|c|}{Sample size in first stage $N^{(1)}$} \\
 \hline
$N=150$: & 15 &  38 & 60 &  83 & 105 & 128 & 150 \\
$N=250$: & 15 &  54 & 93 & 133 & 172 & 211 & 250 \\
\hline
\end{tabular}
\caption{First stage sample size}
\label{tab:timing}
\end{table}

A fixed design is also included for comparison to the adaptive
designs. This corresponds to the scenario when the sample size for the
first stage is 150 or 250 for $N=150$ and $N=250$ respectively. For
the fixed designs, all study participants are assigned to
pre-specified doses according to the starting design and there is no
updating of the design. Two types of updating were considered at the
interim analysis; ML and Bayesian updating. For each of the $448$
scenarios outlined above, $5,000$ trials were simulated.

\subsection{Simulation metrics}

To ensure comparability of the two approaches for design updating we
used ML estimation at study end to estimate the dose-response curve in
both situations. That means also in the situation where Bayesian
updating was used at the interim analysis, at study end ML was used to
estimate the dose-response model parameters. The parameter bounds for
$\theta_3,\theta_4$ as discussed in Section
\ref{sec:maxim-likel-updat} will be used for fitting the dose-response
model.

\subsubsection{Efficiency of the second stage design}
The relative efficiency of the second stage design is defined as
\begin{equation}
\mathrm{RelEff}=\frac{|\mathbf{M}(\xi^{(2)},\boldsymbol{\theta})|}{|\mathbf{M}(\xi^{*},\boldsymbol{\theta})|}
\end{equation}
This is the ratio of the determinant of the information matrix
evaluated at the second stage design $\xi^{(2)}$ to the determinant of
the information matrix evaluated at the locally optimal design
$\xi^{*}$, assuming that the true parameter values
$\boldsymbol{\theta}$ are known. The relative efficiency is bound
between 0 and 1. The closer the relative efficiency is to 1, the
closer the second stage design is to truly optimal.

Whilst the true sigmoid Emax model parameter values are known for the
Emax and the sigmoid Emax model scenario, best fitting ``pseudo-true''
parameter values must be calculated for the linear and quadratic model
scenarios. This is achieved by using least-squares regression to fit
the sigmoid Emax model to the true mean function from the linear and
quadratic dose-response models at the doses $\bm x$. The best fitting
sigmoid Emax model parameters for the linear model were $\bm
\theta=(-0.0396, -4.305, 12, 1.349)$ and for the quadratic model $\bm
\theta=(-0.06617, -1.661, 1.823, 1.948)$.

\subsubsection{Mean absolute error}

As a second metric we will investigate the mean absolute error (MAE)
for estimating the dose-response curve (as in \cite{Bornkamp07}). This
is defined as
\begin{equation}
  \mathrm{MAE}=\frac{1}{q}\sum_{i=1}^q |\eta(x_i,\hat{\boldsymbol{\theta}})-\eta_{true}(x_i)|,
\end{equation}
where $\hat{\boldsymbol{\theta}}$ is the ML estimate at study end, and
$\eta_{true}(x)$ the true dose-response model function in this
scenario.  The mean absolute error is calculated for both the adaptive
($\mathrm{MAE_{adapt}}$) and non-adaptive designs
($\mathrm{MAE_{fixed}}$). For the non-adaptive design the timing of
the `interim' occurs after all $N$ patients have been allocated,
i.e. there is only one stage.

The MAE ratio is defined as
\begin{equation}\label{ratio}
\mathrm{Ratio_{MAE}=\frac{MAE_{fixed}}{MAE_{adapt}}}
\end{equation}
and can be used to compare the adaptive design to the non-adaptive,
with a ratio greater than $1$ suggesting that the adaptive design
performs better.

\section{Results}

\subsection{Efficiency of the second stage design}

\begin{figure}[H]
\begin{center}
\includegraphics[scale=0.75]{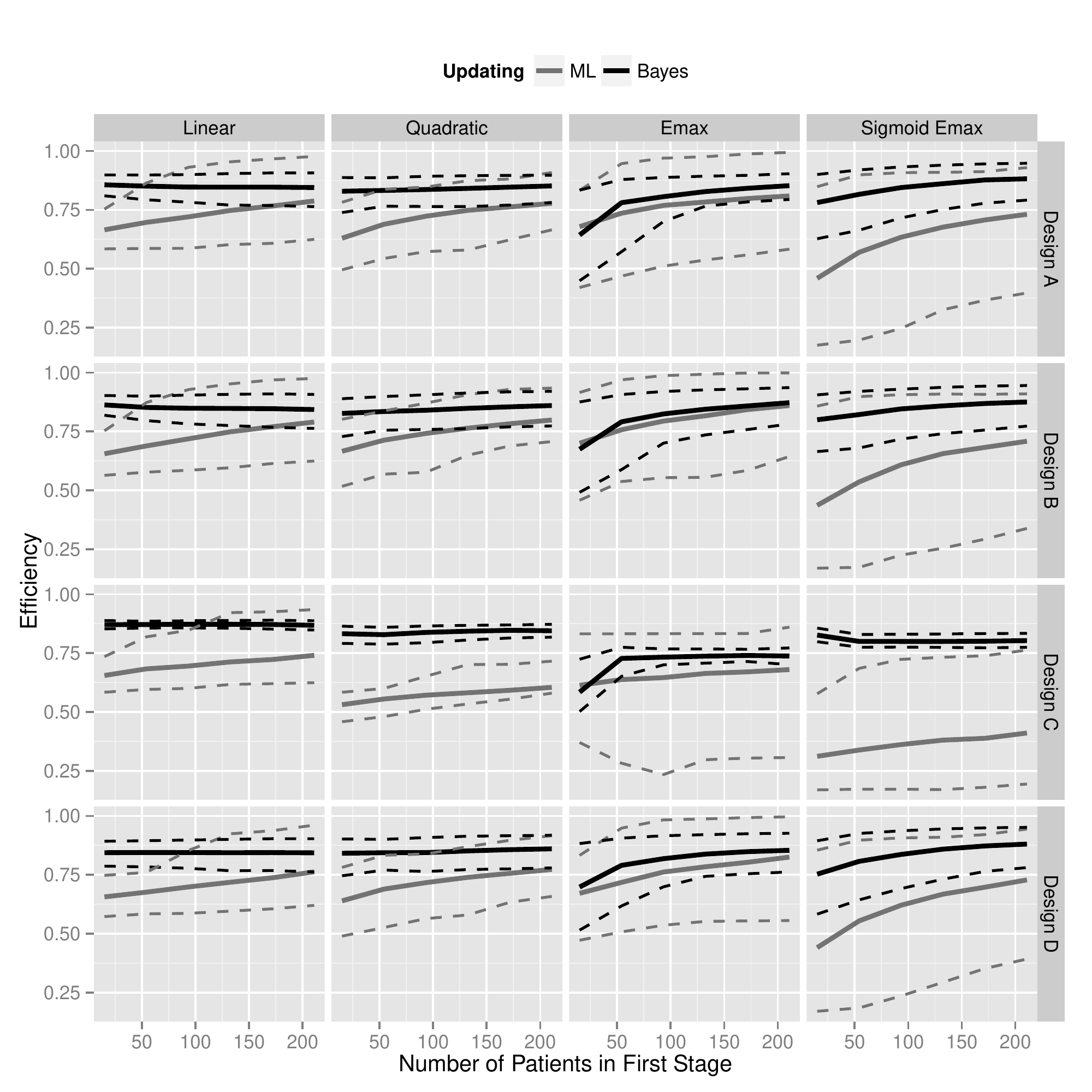}
\end{center}
\caption{Mean (solid line) with 10\% and 90\% quantiles (dashed lines)
  of the observed relative efficiencies of ML and Bayesian
  updating. $N=250$}
\label{fig:eff-mean}
\end{figure}

The distribution of efficiencies of the calculated second stage design
are illustrated in Figure \ref{fig:eff-mean}. One can observe that in
most of the situations the mean efficiency of Bayesian updating
outperforms the ML updating, in particular for early interim analyses,
for the sigmoid Emax model scenario and the ``bad'' starting design C.

For all dose-response scenarios it can be seen that the efficiency
curves for the Bayesian designs only slightly increasing sometimes
almost flat, while for ML-based updating the dependency on the size of
the first stage is much larger, which might be due to the
acknowledgment of parameter uncertainty.

Another interesting observation is that the variability of the
efficiencies for the calculated design are considerably smaller for
Bayesian designs compared to the designs calculated based on ML. So
using a design criterion that acknowledges uncertainty, results in
less variable second stage designs.

The results for $N=150$ are similar and can be found in Appendix
\ref{apdx:results}. In Table \ref{tab:eff} one can observe the
efficiency of the used starting designs, compared to the locally
optimal ones. One can see from the Table that, apart from design C, a
few of the designs have an efficiency of $>0.8$, so that not too much
is gained from the adaptive designs, when comparing the efficiencies
obtained from the adaptive designs in Figure \ref{fig:eff-mean}. In
other words, if the starting design already performs well, then there
is less efficiency to be gained by updating the design at an interim
analysis.

\begin{table}[H]
\centering
\renewcommand{\arraystretch}{1.8}
\begin{tabular}{|c|c|c|c|c|}
\hline
 & Linear & Quadratic & Emax & Sig Emax \\
  \hline
  Design A & 0.91 & 0.61 & 0.62 & 0.73 \\
  Design B & 0.89 & 0.92 & 0.79 & 0.58 \\
  Design C & 0.22 & 0.03 & 0.19 & 0.12 \\
  Design D & 0.81 & 0.76 & 0.63 & 0.86 \\
   \hline
\end{tabular}
\caption{Efficiencies of the different starting designs compared to
  the corresponding local D optimal designs for the sigmoid Emax model}
\label{tab:eff}
\end{table}

\subsection{Mean Absolute Error}

\begin{figure}[H]
\begin{center}
\includegraphics[scale=0.75]{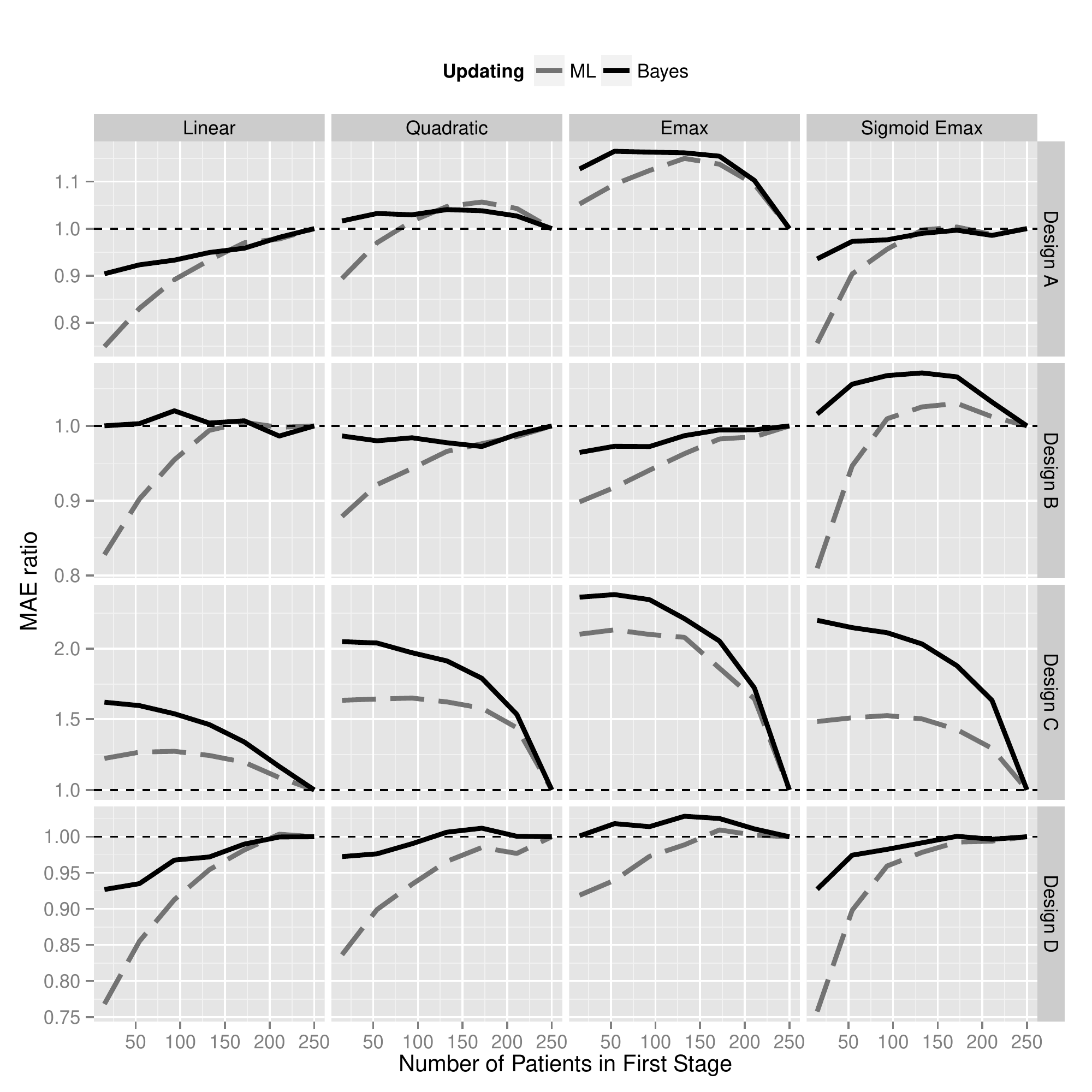}
\end{center}
\caption{Ratio of the MAE: $N=250$}
\label{fig:MAE-N250}
\end{figure}

The ratio of the mean absolute error of the fixed versus the adaptive
design is illustrated for each starting design when the sample size
$N=250$, in Figure \ref{fig:MAE-N250}.

As can be seen in Figure \ref{fig:MAE-N250}, Bayesian updating seems
to outperform ML updating of the design, also in this metric. This
happens in particular for early interim analyses. When the size of the
first stage increases Bayesian and ML updating become more
similar. This is a quite general conclusion across all scenarios.

It can be seen that the benefit of adaption depends strongly on the
starting design. For starting design C one gains in all scenarios,
while in other situations the gain is less apparent. For example, for
starting design A one gains something by adapting in the Emax scenario
and not much for the sigmoid Emax scenario, while for starting design
B it is the other way round. This is also consistent with the
efficiencies of the starting design in Table \ref{tab:eff}, all
scenarios where a benefit of adaption could be observed in Figure
\ref{fig:MAE-N250} correspond to scenarios, where the starting design
has an efficiency of less than 0.7.

The MAE ratio plots corresponding to the smaller sample size of 150
can be found in Appendix \ref{apdx:results}. The findings for $N=150$
are similar to those presented above for $N=250$.

\section{Conclusions}

From the simulation performed it seems that Bayesian updating of
parameter estimates at interim analysis can be used to account for
parameter uncertainty. Assigning functional uniform priors to the
parameters has been shown to increase the efficiency of the second
stage design when compared to ML updating methods, at least in the
scenarios studied. The improvement was particularly pronounced for
early interim analyses. It would be reasonable to assume that if
accurate and valid informative priors could be derived for the
parameters, then the improvement may increase further.

An interesting side-result of the simulations is that the benefit of
including an interim analysis has not been shown to universally to
improve the performance of a dose-finding study. This result is
consistent with other simulation studies \citep{Miller07, Dragalin10,
  Bornkamp11, Jones11}. Factors such as the signal to noise ratio at
interim, or the efficiency of the starting design (as discussed in
\cite{Dette13}) need to be taken into account. For the rather
realistic scenarios discussed in this paper, not much benefit could be
expected, as the starting designs, with exception of design C, all
already performed relatively well. So most benefit is expected for
strongly mis-specified starting designs, such as design C. Practical
experience suggests that such a scenario, where the increasing part of
the dose-response curve is missed in the starting design, is actually
not so rare. So in cases of great uncertainty on the dose-response
curve (at trial design stage) an adaptive design is a valuable option
to consider.

A potentially promising modification of the interim updating strategy,
could be ``update-on-demand'' adaptive designs. Here one would
evaluate the efficiency of the starting design at interim based on
current information (\textit{i.e.} the current posterior distribution,
or ML estimate). One would then only update the design, if the
estimated efficiency of the starting design is smaller than some
threshold. If the estimated efficiency of the starting design is larger
than some threshold one would continue with the starting design until
study end.

\pagebreak

\appendix

\section{K-means approximation of the full posterior distribution}
\label{apdx:kmeans}
\subsection{Justification of k-means approximation of posterior distribution}
Suppose one would like to approximate an integral
\begin{equation}
\int g(\bm \theta)\pi(\bm \theta)d\bm \theta,
\end{equation}
where $\pi(\bm \theta)$ is the posterior density, with $\bm \theta \in \mathbb{R}^d$, and a posterior
sample $\bm \theta_1,\ldots,\bm \theta_T$ is available.

Suppose also that the function $g$ is expensive to evaluate so that the plain Monte
Carlo estimate, $\frac{1}{T}\sum_{t=1}^Tg(\bm \theta_t)$, is not
feasible to evaluate, as $T$ is large.

One approach is to use the k-means clustering algorithm to obtain
cluster centers $\bm a_1, \ldots,\bm a_k \in \mathbb{R}^d$ with $k <<
T$. This approach works by minimizing
\begin{equation}
  \label{eq:kmeans}
  \sum_{i=1}^k \sum_{\bm \theta_t  \in \Omega_i}||\bm a_i-\bm \theta_t||^2,
\end{equation}
with respect to the cluster centers $\bm a_1, \ldots,\bm a_k \in
\mathbb{R}^d$, where $\Omega_i$ is given by $\Omega_i=\{\bm \theta \in
\mathbb{R}^d|\; ||\bm \theta -\bm a_i|| = \underset{i \in \{1, \ldots,
  k\}}{\min} ||\bm \theta -\bm a_i||\}$. That means that every point
$\bm \theta \in \mathbb{R}^d$ belongs to one set $\Omega_i$. The
probability of the sets $\Omega_i$ under $\pi$:
$\alpha_i=\int_{\Omega_i}\pi(\bm \theta)d\bm \theta$ can be estimated
by the proportion $\widehat{\alpha}_i$ of samples within $\Omega_i$.

Using these considerations one can come up with a discrete
approximation of the posterior distribution of form $\tilde{\pi}(\bm
\theta)=\sum_{i=1}^k \widehat{\alpha_i}\delta_{\bm a_i}(\bm \theta)$,
where $\delta_{\bm a}(\bm \theta)$ denotes the function that is equal
to 1 if $\bm \theta = \bm a$ and 0 otherwise.

Now the idea is to approximate the integral $\int g(\bm \theta)\pi(\bm
\theta)d\bm \theta$ by the corresponding integral with respect to
$\tilde{\pi}$, being given by $\sum_{i=1}^k \widehat{\alpha}_ig(\bm
a_i)$.

Now the integration error is given by
\begin{eqnarray}
  \label{eq:trafo}
&&  \left|\int g(\bm \theta)\pi(\bm \theta)d\bm \theta-\sum_{i=1}^k
    \widehat{\alpha}_i g(\bm a_i)\right| \nonumber \\
&=& \left|\sum_{i=1}^k\int_{\Omega_i}g(\bm \theta)\pi(\bm \theta)d\bm
  \theta-\alpha_ig(\bm a_i)+\alpha_ig(\bm a_i)-\widehat{\alpha}_ig(\bm a_i) \right| \nonumber\\
&=& \left|\sum_{i=1}^k\int_{\Omega_i}(g(\bm \theta)-g(\bm a_i))\pi(\bm \theta)d\bm
  \theta+\alpha_ig(\bm a_i)-\widehat{\alpha}_ig(\bm a_i) \right| \nonumber\\
&\leq& \left|\sum_{i=1}^k\int_{\Omega_i}(g(\bm \theta)-g(\bm a_i))\pi(\bm \theta)d\bm
  \theta\right|+\left|(\alpha_i-\widehat{\alpha}_i)g(\bm a_i)\right|
\end{eqnarray}

The term on the right of Equation (\ref{eq:trafo}) will converge
almost surely towards zero when k-means is used to derive $\bm
a_1,\ldots, \bm a_k$, see \cite{Pollard81}. As the number of
samples $T$ will be large in relation to $k$, this term will typically
converge quite quick towards 0.

The term on the left can be treated with arguments as in
\cite[Proposition 7]{Page97}. One can derive an upper bound as
\begin{equation}
  \label{eq:interror}
\sum_{i=1}^k \int_{\Omega_i}|g(\bm a_i)-g(\bm \theta)|\pi(\bm
\theta)d\bm \theta \leq L\sum_{i=1}^k \int_{\Omega_i}||\bm a_i-\bm \theta||\pi(\bm \theta)d\bm \theta
\end{equation}
which holds under the assumption that $g$ is Lipschitz continuous with
bound $L$. Minimizing this upper bound with respect to $\bm a_1,
\ldots,\bm a_k$ will hence minimize an upper bound of the integration
error.  In practice of course \eqref{eq:interror} cannot be optimized
directly as $\pi(\bm \theta)$ is not available in a closed form. Only
a sample $\bm \theta_1,\ldots,\bm \theta_T$ is available. The
empirical version of the upper bound in \eqref{eq:interror} is
proportional to
\begin{equation}
  \label{eq:empirical}
  \sum_{i=1}^k \sum_{\bm \theta_t \in \Omega_i}||\bm a_i-\bm \theta_t||,
\end{equation}
which is very similar to the k-means objective function, see Equation
(\ref{eq:kmeans}). \cite{Graf02} consider the speed of
convergence of the empirical version. But again as $T$ will typically
rather large this will be less of an issue.

\section{Simulation scenario}
\label{apdx:scenario}

The simulation scenario used to investigate the efficiency of the
k-means approximation in terms of relative performance and running
time, used as starting design five unequally spaced doses,
$\xi=(0,1,2,4,8)$ (Design B) with balanced allocations.  The sigmoid
Emax dose-response model was used to simulate the response data with
 \begin{equation}
 \eta(x,\boldsymbol{\theta})=\frac{-1.70x^5}{(4^5+x^5)}
 \end{equation}
 and an independent normally distributed residual error term with mean
 $0$ and variance $4.5$.

\section{Simulation results for $N=150$}
\label{apdx:results}

\begin{figure}[h]
\begin{center}
\includegraphics[scale=0.7]{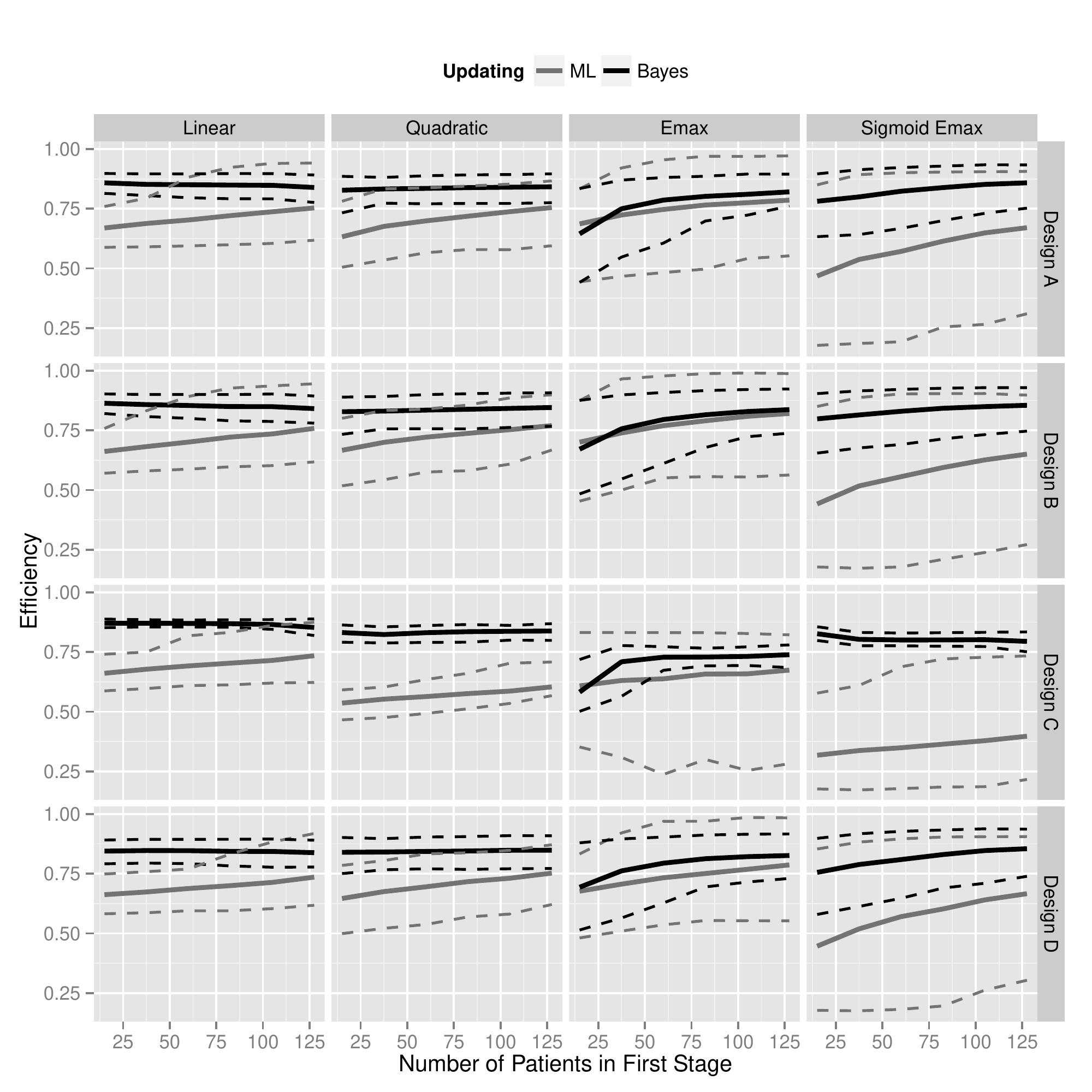}
\end{center}

\caption{Mean (solid line) with 10\% and 90\% quantiles (dashed lines)
  of the observed relative efficiencies of ML and Bayesian
  updating. $N=250$}
\label{fig:MeanEff-N150}
\end{figure}

\begin{figure}[h]
\begin{center}
\includegraphics[scale=0.7]{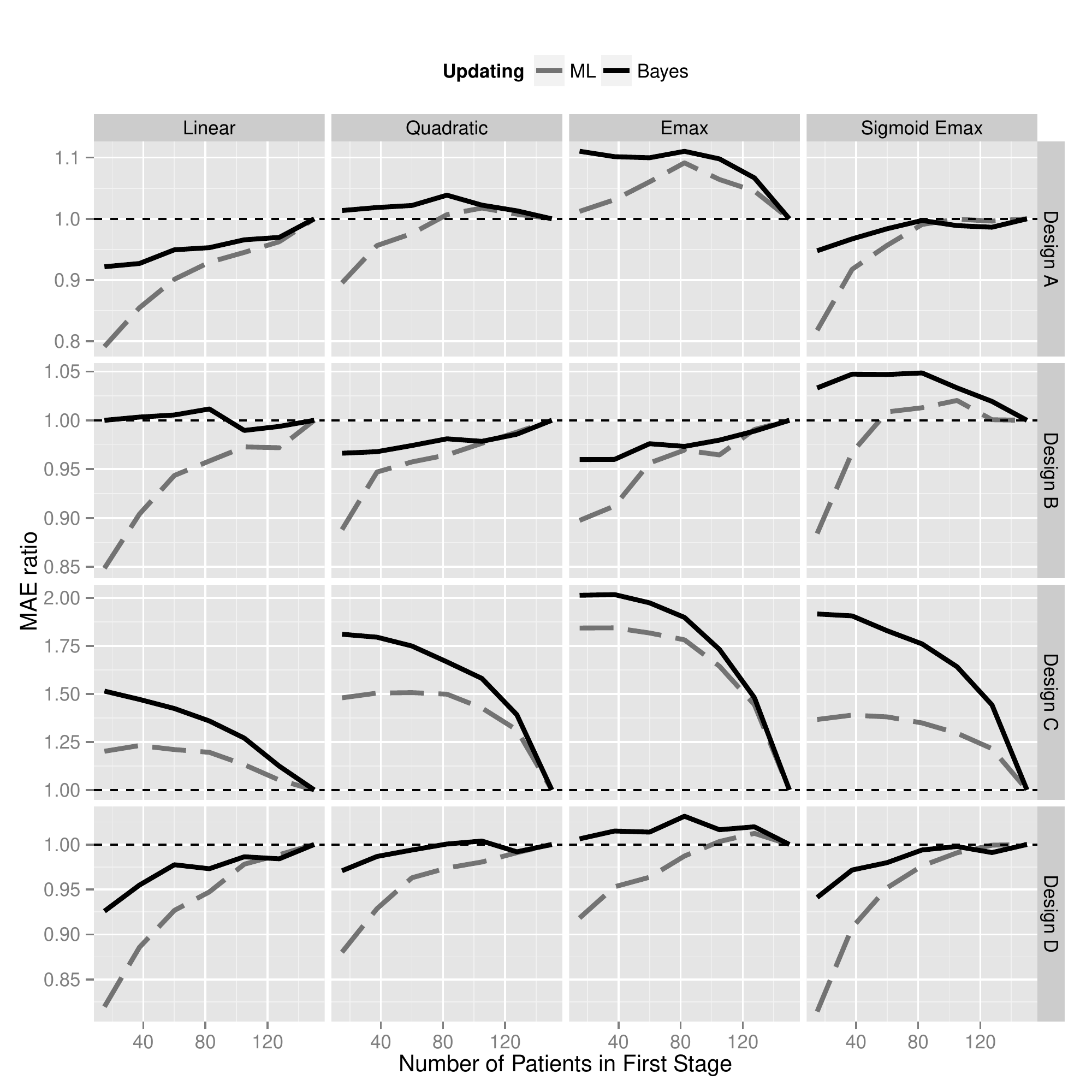}
\end{center}
\caption{Ratio of the MAE: $N=150$}
\label{fig:MAE-N150}
\end{figure}

\clearpage

\bibliographystyle{apa}

\begin{thebibliography}{}

\bibitem[\protect\astroncite{Atkinson et~al.}{2007}]{Atkinson07}
Atkinson, A., Donev, A., and Tobias, R. (2007).
\newblock {\em Optimum Experimental Design, with SAS}.
\newblock Oxford Statistical Science Series. Oxford University Press.

\bibitem[\protect\astroncite{Berry et~al.}{2010}]{Berry10}
Berry, S., Spinelli, W., Littman, G.~S., Liang, J.~Z., Fardipour, P., Berry,
  D.~A., Lewis, R.~J., and Krams, M. (2010).
\newblock A {B}ayesian dose-finding trial with adaptive dose expansion to
  flexibly assess efficacy and safety of an investigational drug.
\newblock {\em Clinical Trials}, 7:121--135.

\bibitem[\protect\astroncite{Bornkamp}{2012}]{Bornkamp12}
Bornkamp, B. (2012).
\newblock Functional uniform priors for nonlinear modeling.
\newblock {\em Biometrics}, 68:893--901.

\bibitem[\protect\astroncite{Bornkamp}{2014}]{Bornkamp14}
Bornkamp, B. (2014).
\newblock Practical considerations for using functional uniform prior
  distributions for dose-response estimation in clinical trials.
\newblock {\em Biometrical Journal}, 0:0--0.

\bibitem[\protect\astroncite{Bornkamp et~al.}{2011}]{Bornkamp11}
Bornkamp, B., Bretz, F., Dette, H., and Pinheiro, J. (2011).
\newblock Response-adaptive dose-finding under model uncertainty.
\newblock {\em The Annals of Applied Statistics}, 5:1611--1631.

\bibitem[\protect\astroncite{Bornkamp et~al.}{2007}]{Bornkamp07}
Bornkamp, B., Bretz, F., Dmitrienko, A., Enas, G., Gaydos, B., Hsu, C., Konig,
  F., Krams, M., Liu, Q., Neuenschwander, B., Parke, T., and Pinheiro, J.
  (2007).
\newblock Innovative approaches for designing and analyzing adaptive
  dose-finding trials.
\newblock {\em Journal of Biopharmaceutical Statistics}, 17:965--995.

\bibitem[\protect\astroncite{Bornkamp et~al.}{2013}]{Bornkamp13}
Bornkamp, B., Pinheiro, J., and Bretz, F. (2013).
\newblock {\em DoseFinding: Planning and Analyzing Dose Finding Experiments}, r
  package version 0.9-6 edition.

\bibitem[\protect\astroncite{Dette et~al.}{2013}]{Dette13}
Dette, H., Bornkamp, B., and Bretz, F. (2013).
\newblock On the efficiency of two-stage response-adaptive designs.
\newblock {\em Statistics in Medicine}, 32:1646--1660.

\bibitem[\protect\astroncite{Dragalin et~al.}{2010}]{Dragalin10}
Dragalin, V., Bornkamp, B., Bretz, F., Miller, F., Padmanabhan, S., Patel, N.,
  Perevozskaya, I., Pinheiro, J., and Smith, J. (2010).
\newblock A simulation study to compare new adaptive dose-ranging designs.
\newblock {\em Statistics in Biopharmaceutical Research}, 2:487--512.

\bibitem[\protect\astroncite{Dragalin et~al.}{2007}]{Dragalin07}
Dragalin, V., Hsuan, F., and Padmanabhan, S. (2007).
\newblock Adaptive designs for dose-finding studies based on sigmoid emax
  model.
\newblock {\em Journal of Biopharmaceutical Statistics}, 17:1051--1070.

\bibitem[\protect\astroncite{Gelman et~al.}{2003}]{Gelman03}
Gelman, A., Carlin, J., Stern, H., and Rubin, D. (2003).
\newblock {\em Bayesian Data Analysis}.
\newblock Chapman \& Hall/CRC.

\bibitem[\protect\astroncite{Ghosh et~al.}{2006}]{Ghosh06}
Ghosh, J., Delampady, M., and Samanta, T. (2006).
\newblock {\em An Introduction to Bayesian Analysis: theory and methods}.
\newblock Springer.

\bibitem[\protect\astroncite{Graf and Luschgy}{2002}]{Graf02}
Graf, S. and Luschgy, H. (2002).
\newblock Rates of convergence for the empirical quantization error.
\newblock {\em Annals of Probability}, 30:874--897.

\bibitem[\protect\astroncite{Grieve and Krams}{2005}]{Grieve05}
Grieve, A.~P. and Krams, M. (2005).
\newblock Astin: a bayesian adaptive dose-response trial in acute stroke.
\newblock {\em Clinical Trials}, 2:340--351.

\bibitem[\protect\astroncite{Hartigan and Wong}{1979}]{Hartigan79}
Hartigan, J.~A. and Wong, M.~A. (1979).
\newblock A k-means clustering algorithm.
\newblock {\em Journal of the Royal Statistical Society: Series C (Applied
  Statistics)}, 28:100--108.

\bibitem[\protect\astroncite{Jennrich}{1969}]{Jennrich1969}
Jennrich, R.~I. (1969).
\newblock Asymptotic properties of nonlinear least squares estimation.
\newblock {\em Annals of Mathematical Statistics}, 40:633--643.

\bibitem[\protect\astroncite{Jones et~al.}{2011}]{Jones11}
Jones, B., Layton, G., Richardson, H., and Thomas, N. (2011).
\newblock Model-based bayesian adaptive dose-finding designs for a phase ii
  trial.
\newblock {\em Statistics in Biopharmaceutical Research}, 3:276--287.

\bibitem[\protect\astroncite{Kirby et~al.}{2011}]{Kirby10}
Kirby, S., Brain, P., and Jones, B. (2011).
\newblock Fitting emax models to clinical trial dose-response data.
\newblock {\em Pharmaceutical Statistics}, 10(2):143--149.

\bibitem[\protect\astroncite{MacDougall}{2006}]{MacDougall06}
MacDougall, J. (2006).
\newblock {\em Dose Finding in Drug Development}, chapter Analysis of
  Dose-Response Studies - Emax Model, pages 127--145.
\newblock Statistics for Biology and Health. Springer.

\bibitem[\protect\astroncite{Miller et~al.}{2007}]{Miller07}
Miller, F., Guilbaud, O., and Dette, H. (2007).
\newblock Optimal designs for estimating the interesting part of the
  dose-effect curve.
\newblock {\em Journal of Biopharmaceutical Statistics}, 17:1097--1115.

\bibitem[\protect\astroncite{Pag\`{e}s}{1997}]{Page97}
Pag\`{e}s, G. (1997).
\newblock A space quantization method for numerical integration.
\newblock {\em Journal of Computational and Applied Mathematics}, 89:1--38.

\bibitem[\protect\astroncite{Pollard}{1981}]{Pollard81}
Pollard, D. (1981).
\newblock Strong consistency of $k$-means clustering.
\newblock {\em Annals of Statistics}, 9:135--140.

\bibitem[\protect\astroncite{Pukelsheim}{2006}]{Pukelsheim06}
Pukelsheim, F. (2006).
\newblock {\em Optimal Design of Experiments}.
\newblock SIAM.

\bibitem[\protect\astroncite{Temple}{2012}]{Temple12}
Temple, J. (2012).
\newblock {\em Adaptive Designs for Dose-Finding Trials}.
\newblock Thesis (doctor of philosophy (phd)), University of Bath.

\bibitem[\protect\astroncite{Thomas}{2006}]{Thomas06}
Thomas, N. (2006).
\newblock Hypothesis testing and bayesian estimation using a sigmoid emax model
  applied to sparse dose-response designs.
\newblock {\em Journal of Biopharmaceutical Statistics}, 16:657--677.

\bibitem[\protect\astroncite{Wu}{2012}]{Wu12}
Wu, J. (2012).
\newblock {\em Advances in k-means clustering: a data mining thinking}.
\newblock Springer.

\end{thebibliography}

\end{document}